# Mid-wave and Long-wave IR Linear Dichroism in a Hexagonal Perovskite Chalcogenide

Shanyuan Niu[†], Huan Zhao[§], Yucheng Zhou[†], Huaixun Huyan[†], Boyang Zhao[†], Jiangbin Wu[§], Stephen B. Cronin[§], Han Wang[§], Jayakanth Ravichandran*[†§]

[†]Mork Family Department of Chemical Engineering and Materials Science, University of Southern California, Los Angeles, CA 90089, USA

[§]Ming Hsieh Department of Electrical Engineering, University of Southern California, Los Angeles, CA 90089, USA

**ABSTRACT:** Mid-wave infrared (IR) and long-wave IR spectral ranges are of growing interest in various applications such as thermal imaging, thermography-based remote sensing, and night vision. Materials widely used for IR photodetectors in this regime include cadmium mercury telluride alloys and nanostructures of compound semiconductor. The materials development for IR optics will drive down the cost of IR optical systems and enable larger scale deployment. Here, we report a mid-wave IR responsive material composed of earth abundant and non-toxic elements, $Sr_{1+x}TiS_3$. It has a highly anisotropic quasi-one-dimensional structure similar to hexagonal perovskites. We grew large, high quality single crystals and studied its anisotropic optical properties. We observed two distinct optical absorption edges at ~2.5 μm and ~5 μm, respectively, for linear polarizations along two principle axes. The material demonstrated strong and broadband linear dichroism spanning mid-wave IR and long-wave IR, with a dichroitic ratio of up to 22.

Mid-wave IR (MWIR) and long-wave IR (LWIR) spectral ranges cover the second (~3-5 μm) and the third (~8-14 μm) atmospheric transmission windows in the infrared radiation.[1] Hence, MWIR and LWIR imaging or thermography systems have broad scientific, industrial, and military applications such as molecular fingerprint imaging, remote sensing, free space telecommunication, target discrimination, and surveillance.[2-6] Conventional bulk semiconductors such as Si and III-V alloys do not interact with the low energy photons effectively as one approaches the MWIR. The widely used materials in this regime include HgCdTe cadmium mercury telluride alloys,[7,8] and quantum-well or quantum-dot nanostructures.[9-12] Several issues such as their environmentally hazardous composition and the need for sophisticated growth and fabrication processes post serious challenges to drive down the cost of the MWIR and LWIR detection devices. Besides the materials for detectors, very few materials possess any appreciable anisotropy and/or non-linearity to act as linear or non-linear optical elements in this regime, which is another key to this puzzle. Hence, polarization-sensitive interaction with MWIR and LWIR light heavily relies on narrow bandwidth, extrinsic geometric patterns, which adds another layer of limitations to the performance of these systems. Thus, developing new broadband, MWIR and LWIR responsive materials and identifying intrinsic anisotropic functionalities of IR responsive materials remains an important challenge.

Linear dichroism is a property of a material, which represents the difference in the attenuation of light with polarizations parallel or perpendicular to a crystallographic axis. The spectroscopic technique that probes this property goes by the same name.[13] This intrinsic anisotropic response in materials is key to enable novel photonic devices such as polarization rotation, polarizing filters, light modulators, and polarization sensitive photodetectors.[13-16] Previous attempts to demonstrate such an effect in technologically relevant MWIR and LWIR regime have mainly focused on the one-dimensional nanostructures, but the aspect ratios make the scalable, device fabrication and alignment of the nanoscale features challenging. Recent studies have shown linear dichroism in two-dimensional layered materials such as GeSe[13] in SWIR and black phosphorus (b-P)[17] in MWIR, which arises from their in-plane structural anisotropy.[18,19] Within a few years of development, various optoelectronic devices based on b-P, such as high gain mid-IR photodetectors,[20] polarization-sensitive photodetectors[21] have been demonstrated. Very recently, we have demonstrated more pronounced in-plane anisotropy in the MWIR and LWIR spectral ranges in a ternary hexagonal perovskite chalcogenide $BaTiS_3$ (BTS), where parallel quasi-one-dimensional (quasi-1D) chains are formed by face-sharing $TiS_6$ octahedra. We reported giant birefringence in the MWIR and LWIR region for this material due to the large structural and chemical anisotropy between intra- and inter-chain directions.[22]

Here, we report strong and broadband linear dichroism spanning MWIR and LWIR in another hexagonal perovskite with a quasi-1D structure, $Sr_{1+x}TiS_3$ (STS). Large, high quality single crystals were grown using chemical vapor transport method. We used x-ray diffraction and vibrational spectroscopy to establish the structural anisotropy of the material i.e. the directions along and across the chains. We observed two distinctive optical edges at ~2.5 μm and ~5 μm for linear polarizations parallel and perpendicular to the chains respectively. Strongest linear dichroism appeared in MWIR with a dichroitic ratio of well over 20.

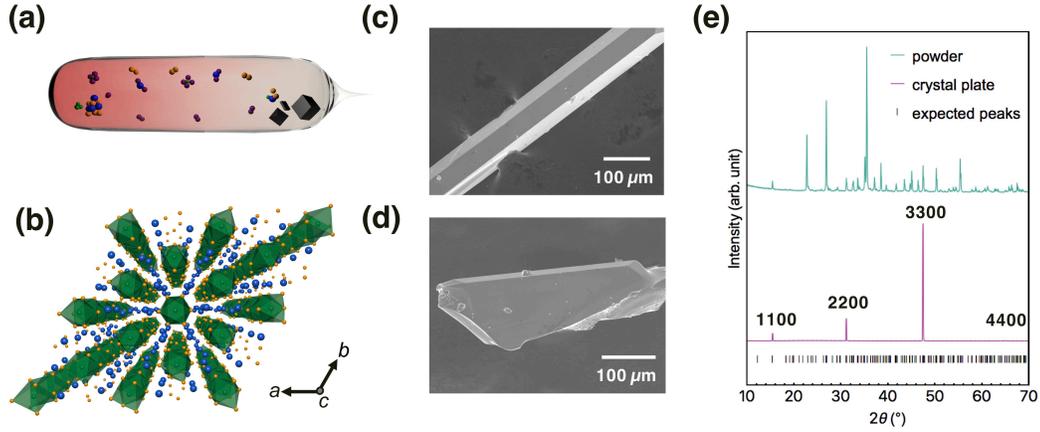

Figure 1. (a) The schematic for crystal growth. (b) The schematic for crystal structure. The SEM images of a crystal wire (c) and a crystal platelet (d). (e) Overlay of powder XRD, out-of-plane XRD scan of a crystal platelet, and expected peaks.

We grew single crystals of STS using chemical vapor transport method with iodine as a transporting agent. The synthetic procedure is similar to that used for other perovskite sulfides reported earlier.[22,23] The schematic of crystal growth is shown in **Figure 1(a)**. The structure of STS falls in the broad category of $BaNiO_3$-related structure,[24] with face-sharing $TiS_6$ octahedral chains and Sr chains extending along $c$-axis and aligned in a hexagonal setting of trigonal symmetry (**Figure 1(b)**).[25,26] However, the structure is distorted to a more complex incommensurate modulated structure that cannot be indexed in the three-dimensional formalism.[27,28] $BaNiO_3$ structure has matched periodicities of Ba chains and $NiO_6$ octahedral chains, and thus equivalent number of alkaline metal and transition metal atom. In contrast, the periodicities of the Sr chains and $TiS_6$ chains in the distorted structure are not matched, resulting in an actual composition of $Sr_{1+x}TiS_3$.[27,28] It can be viewed as interpenetration of two sub-cells having common lattice constants perpendicular to the $c$-axis and two different ones along $c$-axis. Previous studies on STS resolved its structure through Rietveld refinement with a (3+1)-dimensional super space group formalism ($h\ k\ l\ m$),[29] where the last two indices refer to the different $c$-axis periodicities for Sr and $TiS_6$ chains.[25,28]

As-grown crystals revealed two predominant morphologies, wires and platelets, as shown in the scanning electron microscope (SEM) images (Figure 1(c), 1(d)). Both types of crystals tend to have smooth and clean naturally terminated facets, presumably corresponding to crystallographic planes. We performed powder X-ray diffraction (XRD) studies on the ground crystals and high resolution out-of-plane scan on the crystal platelet. The out-of-plane scan on the platelet showed one set of sharp Bragg reflections with a d-spacing of 5.74, 2.87, 1.91, and 1.43 Å, respectively. Both the powder pattern and the extracted d-spacing agree with an STS structure with $x=0.145$ reported earlier based on polycrystalline samples.[28] The powder pattern, Bragg reflections from the crystal platelet, and expected peak positions are overlaid in **Figure 1(e)**. The platelet facet was indexed to be along {$1\bar{1}00$}, following the aforementioned (3+1)-dimensional formalism. This confirms that the naturally terminated surface is not the basal plane, but the prismatic plane instead, which enables the large anisotropy between intra- and inter- chain directions to be easily accessed by probing top surface of the platelet. Chemical composition characterization was carried out with energy dispersive analytical X-ray spectroscopy (EDS). EDS measurements on the crystal surface showed only expected elements with minimal signal from oxygen. A representative EDS spectrum at 400X magnification is shown in **Figure 2(a)**. EDS mapping of Sr, Ti, and S elements on a wire and a platelet was shown in **Figure 2(b)**. A consistent ratio of Sr: Ti: S at varied locations and magnifications was obtained to be around 1.1: 1: 2.9. This agrees well with the expected off-stoichiometry in this material.

In order to perform polarization-resolved studies, we need to identify the $c$-axis of the crystal. It is typically readily indicated by the long and sharp cleavage of the crystals, as shown in the optical image of a crystal platelet in **Figure 2(c)**. It is further confirmed by extending the thin film out of plane diffraction to identify the 3-fold rotation axis of the trigonal symmetry. Consecutive out of plane scans were taken in a thin film diffractometer as the crystal plate is tilted with respect to the supposed $c$-axis. An illustrative schematic for the process is shown in **Figure 2(d)**. An XRD map of $2\theta/\omega$ scans at each tilting angle $\psi$ was then constructed (**Figure 2(e)**). It is essentially a map of reciprocal space that is perpendicular to the tilting axis. The reciprocal lattice points are extended to long streaks along $\psi$ due to the relatively poor coherency of the probe in this direction. All the reflections are indexed with aforementioned four-dimensional formalism. All the reflections are ($h\ k\ 0\ 0$) type, confirming that the tilting axis is indeed $c$-axis. We can see expected 5200, 4100, 3000, $5\bar{1}00$, $7\bar{2}00$, and $2\bar{1}00$ reflections, which follow the reflection conditions of $-h+k=3n$, when the crystal is tilted 13.9°, 19.1°, 30°, 40.9°, 46.1°, and 60° away from $1\bar{1}00$ facet. This agrees well with the trigonal crystal symmetry and shows high quality of the crystal.

We performed the polarization-resolved Raman spectroscopy to study the vibrational response of such an anisotropic

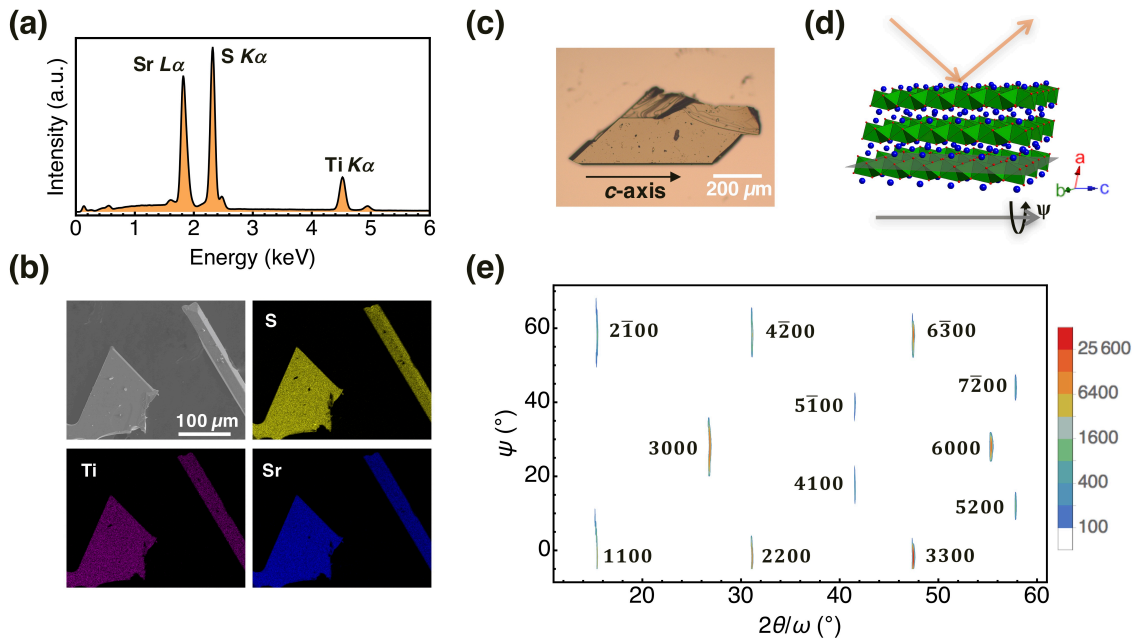

Figure 2. (a) EDS spectrum of the crystal. (b) SEM image and EDS mapping of S, Ti, and Sr elements of a crystal wire and a platelet. (c) Optical image of a crystal plate. (d) The schematic for rotational XRD mapping process, consecutive $2\theta/\omega$ out-of-plane XRD scans are taken on the crystal plate as a function of its tilting angle $\psi$. (e) The rotational XRD map of a crystal plate. Intensity of the reflections are indicated by the color bar contour.

structure. Back-scattering geometry with a 532 nm laser excitation was used for the measurements.

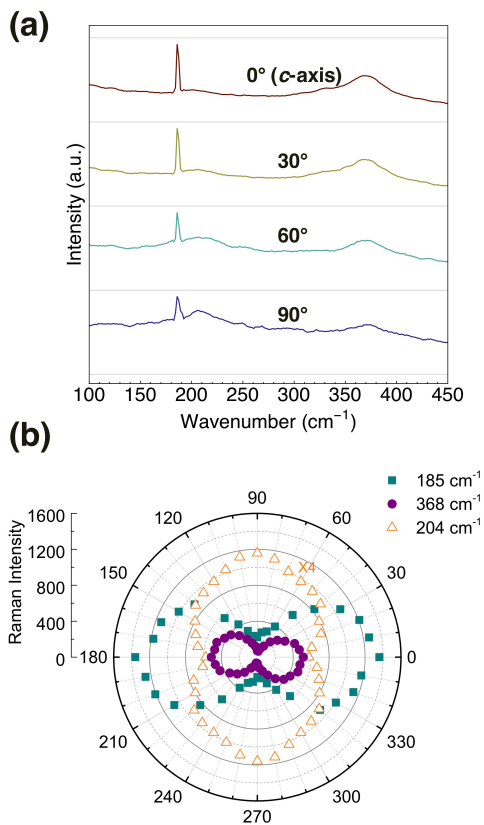

Figure 3. (a) Normalized Raman spectra with incident laser linearly polarized parallel, 30º, 60º, and 90º to the *c*-axis. (b) A polar plot of the three Raman peak intensities as a function of the angle between incident laser polarization and the *c*-axis.

We recorded several sets of measurements with the polarization of the laser rotated in steps of 10º at room temperature. We have observed three Raman peaks at 185, 204, and 368 cm$^{-1}$, respectively. Peaks at 185 and 368 cm$^{-1}$ show the highest intensity for the incident polarization parallel to *c*-axis and the weakest for the perpendicular polarization, while peak at 204 cm$^{-1}$ shows the opposite trend. Raman spectra with polarization parallel, 30º, 60º, and 90º to the *c*-axis is overlaid in **Figure 3(a)**. Intensity of the three peaks in all 18 sets of measurements were plotted in a polar coordinate (**Figure 3(b)**). This large anisotropy in Raman peak intensities can also be used to identify the *c*-axis of the crystal platelet.

Polarization-resolved infrared spectroscopy was performed to study the optical response in the infrared region. Normal incidence transmission measurements were carried out for controlled polarization with 10º steps on the prismatic plane of a crystal platelet. The *c*-axis direction on the facet was predetermined with rotational XRD map and Raman measurements discussed earlier. When the incident polarization is parallel to the *c*-axis, we observed a sharp optical absorption edge at ~2.5 μm (0.5 eV). However, when the polarization is perpendicular to the c-axis, the transmission shows a different sharp absorption feature at ~5 μm (0.25 eV) instead. The crystal platelet appears highly absorptive for linear perpendicular polarization but remains transparent for the linear parallel polarization in the MWIR region. The strong, broadband dichroic window extends to LWIR region while the dichroic magnitude gradually decreases for longer wavelengths. The absorbance value of the transmission measurements with polarization parallel, 20º, 40º, 60º, 80º, and 90º to the *c*-axis are plotted in **Figure 4(a)**. The transmission and absorbance values at 5 μm for all measurements are shown in a polar plot (**Figure 4(b)**), with a strong dichroic ratio (the ratio between transmission of parallel and perpendicular polarizations) of over 20. Note that at low wavelengths, the accurate determination of the nearly zero transmiss-ion of perpendicular polarization is limited by the instrument, thus leading to an underestimated dichroic ratio. The measured dichroic and absorbance ratios for various polarizations are listed in **Table 1**.

**Table 1. Transmission of various linear polarizations, dichroic ratios, and absorbance ratios.**

| wavelength | 0º (*c*-axis) | 30º | 60º | 90º | dichroic ratio ($I_c/I_a$) | Absorbance ratio ($\alpha_a / \alpha_c$) |
|---|---|---|---|---|---|---|
| 3 μm | 17.8 | 14.6 | 6.5 | 1.8 | 9.9 | 2.3 |
| 4 μm | 46.4 | 37.8 | 14.9 | 2.1 | 22.1 | 5.1 |
| 5 μm | 63.4 | 51.7 | 20.7 | 3.1 | 20.5 | 7.5 |
| 6 μm | 65.1 | 54.3 | 25.4 | 9.0 | 7.2 | 5.6 |
| 7 μm | 70.2 | 60.1 | 32.5 | 16.8 | 4.18 | 5.1 |
| 8 μm | 72.5 | 62.9 | 37.4 | 23.2 | 3.1 | 4.5 |
| 9 um | 73.2 | 63.9 | 39.1 | 25.3 | 2.9 | 4.6 |
| 10 um | 73.8 | 64.6 | 40.2 | 26.5 | 2.8 | 4.5 |

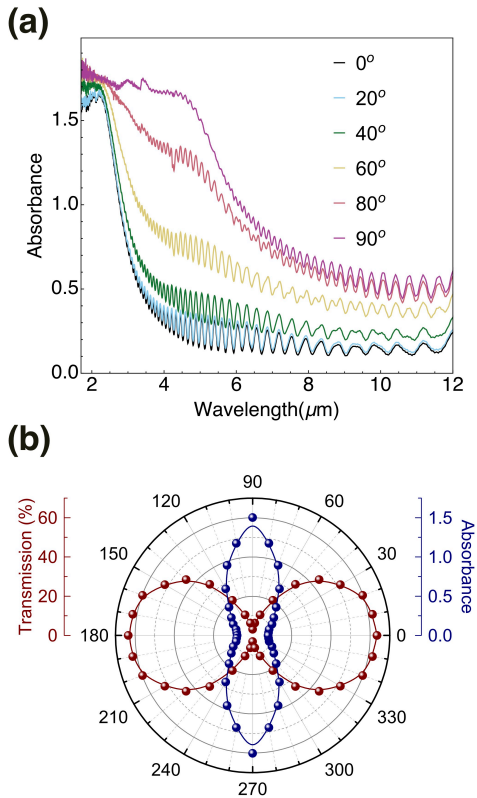

Figure 4. (a) The absorbance for normal incidence with linear polarizations parallel, 20º, 40º, 60º, 80º, and 90º to the *c*-axis in the crystal plate. (b) Transmission and absorbance value plotted in polar coordinate as a function of the incident polarization with respect to *c*-axis.

In conclusion, we have rationally designed and experimentally demonstrated strong and broadband dichroism in a hexagonal perovskite chalcogenide crystal, $Sr_{1+x}TiS_3$, with a quasi-1D structure. Large, high quality single crystals were grown and characterized using structural, chemical and optical probes. X-ray diffraction and anisotropic vibrational response was used to identify the principle axis in the naturally terminated prismatic plane of the crystals. Two distinct optical edges were observed at 2.5 μm and 5 μm for light with linear polarization parallel or perpendicular to the principle axis, respectively. Strong dichroism window spans from 2.5 μm to the longest measured 12 μm spectral range, with the strongest dichroism peaking in the MWIR region reaching a dichroic ratio well over 20. Our study introduces another broadband dichroic material for MWIR/LWIR optics and further reveals the potential of quasi-1D hexagonal perovskite chalcogenides for integrated, polarization-sensitive IR optical systems.

## ASSOCIATED CONTENT

## AUTHOR INFORMATION

**Corresponding Author**

* Jayakanth Ravichandran: jayakanr@usc.edu

**Author Contributions**

‡Shanyuan Niu and Huan Zhao contributed equally.

## ACKNOWLEDGMENT

J.R. and S.N. acknowledge USC Viterbi School of Engineering Startup Funds and support from the Air Force Office of Scientific Research under award number FA9550-16-1-0335. S.N. acknowledges Link Foundation Energy Fellowship. H.W. and H.Z. acknowledges support from the Army Research Office (Grant No. W911NF-16-1-0435) and National Science Foundation (Grant No. ECCS-1653870). We acknowledge the use of facilities at Center for Excellence in Microscopy and MicroAnalysis (CEMMA).

## REFERENCES

(1)  Rogalski, A.; Chrzanowski, K. Infrared Devices and Techniques. *Opto-Electron. Rev.* **2002**, *10*, 111–136.

(2) Petersen, C. R.; Moller, U.; Kubat, I.; Zhou, B.; Dupont, S.; Ramsay, J.; Benson, T.; Sujecki, S.; Abdel-Moneim, N.; Tang, Z.; *et al.* Mid-Infrared Supercontinuum Covering the 1.4-13.3 μm Molecular Fingerprint Region Using Ultra-High NA Chalcogenide Step-Index Fibre. *Nat. Photon.* **2014**, *8*, 830–834.

(3) Seddon, A. B. A Prospective for New Mid-Infrared Medical Endoscopy Using Chalcogenide Glasses. *Int. J. Appl. Glass Sci.* **2011**, *2*, 177–191.

(4) Wartewig, S.; Neubert, R. H. H. Pharmaceutical Applications of Mid-IR and Raman Spectroscopy. *Adv. Drug Deliv. Rev.* **2005**, *57*, 1144–1170.

(5) Hackwell, J. A.; Warren, D. W.; Bongiovi, R. P.; Hansel, S. J.; Hayhurst, T. L.; Mabry, D. J.; Sivjee, M. G.; Skinner, J. W. LWIR/MWIR Imaging Hyperspectral Sensor for Airborne and Ground-Based Remote Sensing. In; Descour, M. R.; Mooney, J. M., Eds.; SPIE, 1996; Vol. 2819, pp. 102–107.

(6) Willer, U.; Saraji, M.; Khorsandi, A.; Geiser, P.; Schade, W. Near- and Mid-Infrared Laser Monitoring of Industrial Processes, Environment and Security Applications. *Opt. Lasers Eng.* **2006**, *44*, 699–710.

(7) Rogalski, A. Infrared Detectors: Status and Trends. *Prog. in Quant. Electron.* **2003**, *27*, 59–210.

(8) Rogalski, A. Toward Third Generation HgCdTe Infrared Detectors. *J. Alloy Compd.* **2004**, *371*, 53–57.

(9) Chow, D. H.; Miles, R. H.; Schulman, J. N.; Collins, D. A.; McGill, T. C. Type II Superlattices for Infrared Detectors and Devices. *Semicond. Sci. Technol.* **1991**, *6*, C47–C51.

(10) Levine, B. F. Quantum-Well Infrared Photodetectors. *J. Appl. Phys.* **1998**, *74*, R1–R81.

(11) Schneider, H.; Liu, H. C. Quantum Well Infrared Photodetectors; Springer, 2006.

(12) Lhuillier, E.; Keuleyan, S.; Liu, H.; Guyot-Sionnest, P. Mid-IR Colloidal Nanocrystals. *Chem. Mater.* **2013**, *25*, 1272–1282.

(13) Wang, X.; Li, Y.; Le Huang; Jiang, X.-W.; Jiang, L.; Dong, H.; Wei, Z.; Li, J.; Hu, W. Short-Wave Near-Infrared Linear Dichroism of Two-Dimensional Germanium Selenide. *J. Am. Chem. Soc.* **2017**, *139*, 14976–14982.

(14) Shen, H.; Wraback, M.; Pamulapati, J.; Dutta, M.; Newman, P. G.; Ballato, A.; Lu, Y. Normal Incidence High Contrast Multiple Quantum Well Light Modulator Based on Polarization Rotation. *Appl. Phys. Lett.* **1993**, *62*, 2908–2910.

(15) Misra, P.; Sun, Y. J.; Brandt, O.; Grahn, H. T. In-Plane Polarization Anisotropy and Polarization Rotation for M-Plane GaN Films on LiAlO2. *Appl. Phys. Lett.* **2003**, *83*, 4327–4329.

(16) Wraback, M.; Shen, H.; Liang, S.; Gorla, C. R.; Lu, Y. High Contrast, Ultrafast Optically Addressed Ultraviolet Light Modulator Based Upon Optical Anisotropy in ZnO Films Grown on R-Plane Sapphire. *Appl. Phys. Lett.* **1999**, *74*, 507–509.

(17) Long, M.; Gao, A.; Wang, P.; Xia, H.; Ott, C.; Pan, C.; Fu, Y.; Liu, E.; Chen, X.; Lu, W.; *et al.* Room Temperature High-Detectivity Mid-Infrared Photodetectors Based on Black Arsenic Phosphorus. *Science Advances* **2017**, *3*, e1700589.

(18) Qiao, J.; Kong, X.; Hu, Z.-X.; Yang, F.; Ji, W. High-Mobility Transport Anisotropy and Linear Dichroism in Few-Layer Black Phosphorus. *Nat. Commun.* **2014**, *5*, 1–7.

(19) Xia, F.; Wang, H.; Jia, Y. Rediscovering Black Phosphorus as an Anisotropic Layered Material for Optoelectronics and Electronics. *Nat. Commun.* **2014**, *5*, 4458.

(20) Yuan, H.; Liu, X.; Afshinmanesh, F.; Li, W.; Xu, G.; Sun, J.; Lian, B.; Curto, A. G.; Ye, G.; Hikita, Y.; *et al.* Polarization-Sensitive Broadband Photodetector Using a Black Phosphorus Vertical P-N Junction. *Nat. Nanotech.* **2015**, *10*, 707–713.

(21) Guo, Q.; Pospischil, A.; Bhuiyan, M.; Jiang, H.; Tian, H.; Farmer, D.; Deng, B.; Li, C.; Han, S.-J.; Wang, H.; *et al.* Black Phosphorus Mid-Infrared Photodetectors with High Gain. *Nano Lett.* **2016**, *16*, 4648–4655.

(22) Niu, S.; Joe, G.; Zhao, H.; Zhou, Y.; Orvis, T.; Huyan, H.; Salman, J.; Mahalingam, K.; Urwin, B.; Wu, J.; *et al.* Giant Optical Anisotropy in a Quasi-One- Dimensional Crystal. *Nat. Photon.* in press **2018**. DOI:10.1038/s41566-018-0189-1

(23) Niu, S.; Huyan, H.; Liu, Y.; Yeung, M.; Ye, K.; Blankemeier, L.; Orvis, T.; Sarkar, D.; Singh, D. J.; Kapadia, R.; *et al.* Bandgap Control via Structural and Chemical Tuning of Transition Metal Perovskite Chalcogenides. *Adv. Mater.* **2017**, *29*, 1604733.

(24) Tranchitella, L. J.; Fettinger, J. C.; Dorhout, P. K.; Van Calcar, P. M.; Eichhorn, B. W. Commensurate Columnar Composite Compounds: Synthesis and Structure of $Ba_{15}Zr_{14}Se_{42}$ And $Sr_{21}Ti_{19}Se_{57}$. *J. Am. Chem. Soc.* **1998**, *120*, 7639–7640.

(25) Gourdon, O.; Petricek, V.; Evain, M. A New Structure Type in the Hexagonal Perovskite Family; Structure Determination of the Modulated Misfit Compound $Sr_{9/8}TiS_3$. *Acta Cryst.* **2000**, *B56*, 409–418.

(26) Gourdon, O.; Jeanneau, E.; Evain, M.; Jobic, S.; Brec, R.; Koo, H. J.; Whangbo, M. H. Influence of the Metal–Metal Sigma Bonding on the Structures and Physical Properties of the Hexagonal Perovskite-Type Sulfides $Sr_{9/8}TiS_3$, $Sr_{8/7}TiS_3$, And $Sr_{8/7}[Ti_{6/7}Fe_{1/7}]S_3$. *J. Solid State Chem.* **2001**, *162*, 103–112.

(27) Saeki, M.; Onoda, M. Preparation of a New Strontium Titanium Sulfide $Sr_xTiS_3$ (x = 1.05 - 1.22) with Infinitely Adaptive Structures. *J. Solid State Chem.* **1993**, *102*, 100–105.

(28) Onoda, M.; Saeki, M.; Yamamoto, A.; Kato, K. Structure Refinement of the Incommensurate Composite Crystal $Sr_{1.145}TiS_3$ Through the Rietveld Analysis Process. *Acta Cryst.* **1993**, *B49*, 929–936.

(29) Van Smaalen, S. Incommensurate Crystal Structures. *Crystallogr. Rev.* **1995**, *4*, 79–202.

Insert Table of Contents artwork here

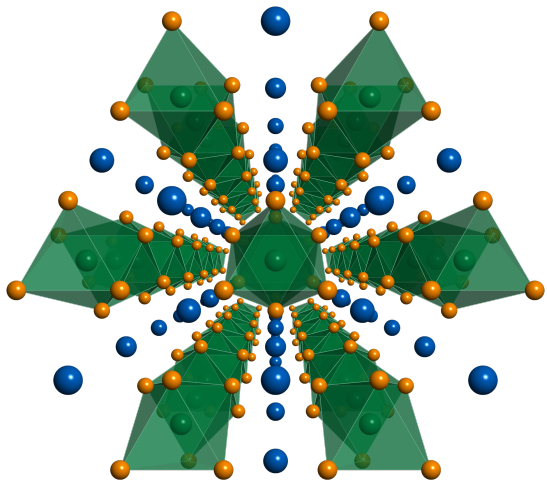 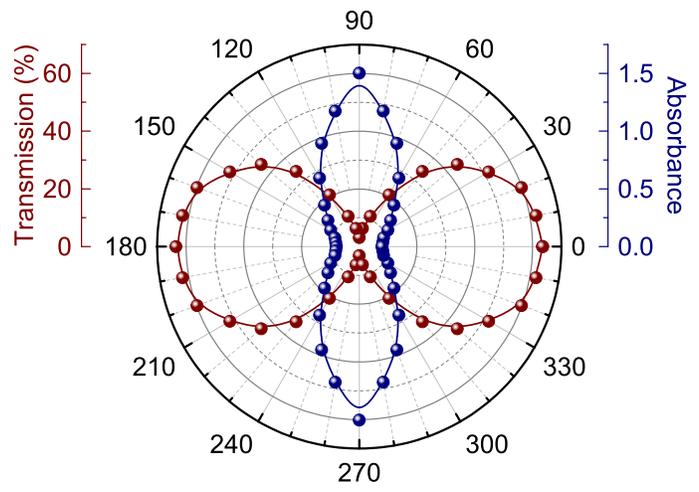